\documentclass[twocolumn]{Trivedi_Papers}
\usepackage{graphicx}
\usepackage[]{cite}
\usepackage{floatrow}
\usepackage{sidecap}
\usepackage{tcolorbox}
\usepackage{notes2bib}
\usepackage{lipsum}
\usepackage[hidelinks]{hyperref}
\usepackage{sidecap}
\usepackage[switch]{lineno}

\usepackage{zref-xr}
\zxrsetup{%
  tozreflabel=false, % not needed, since `zref` is not used otherwise
  toltxlabel=true, % classical LaTeX labels that can be referenced with \ref, \pageref
}

\usepackage{color}
\newcommand{\red}[1]{\textcolor{red}{#1}}
\newcommand{\green}[1]{\textcolor{cyan}{#1}}

\makeatletter
\def\blfootnote{\gdef\@thefnmark{}\@footnotetext}
\makeatother

\newcommand{\blist}{
 \begin{list}{$\bullet$}
  { \setlength{\itemsep}{2pt}
     \setlength{\parsep}{0pt}
     \setlength{\topsep}{6pt}
     } }
\newcommand{\elist}{
  \end{list}  }

\begin{document}

\title{Studying evolution of the primary body axis \textit{in vivo} and \textit{in vitro}}
\vspace{-8pt}
\author{{\authorfont\authorcolor
Kerim Anlas$^{1}$, Vikas Trivedi$^{1,2,*}$}\\[6pt]
%\affilfont
%\affilfont
%\bigskip
%{\sectioncolor\hrule height .6pt}
\medskip
\begin{center}
    \begin{minipage}{0.8\textwidth}
    %\xabstractfont
\bfseries\sffamily{The metazoan body plan is established during early embryogenesis via collective cell rearrangements and evolutionarily conserved gene networks, as part of a process commonly referred to as gastrulation. While substantial progress has been achieved in terms of characterizing the embryonic development of several model organisms, underlying principles of many early patterning processes nevertheless remain enigmatic. Despite the diversity of (pre-)gastrulating embryo and adult body shapes across the animal kingdom, the body axes, which are arguably the most fundamental features, generally remain identical between phyla. Recently there has been a renewed appreciation of \textit{ex vivo} and \textsl{in vitro} embryo-like systems to model early embryonic patterning events. Here, we briefly review key examples and propose that similarities in morphogenesis as well as associated gene expression dynamics may reveal an evolutionarily conserved developmental mode as well as provide further insights into the role of external or extraembryonic cues in shaping the early embryo. In summary, we argue that embryo-like systems can be employed to inform previously uncharted aspects of animal body plan evolution as well as associated patterning rules.}\\
%%\\One-sentence summary 
           \end{minipage}
\end{center}
\vspace{-8pt}
%\medskip
%{\sectioncolor\hrule height .6pt}
}

\maketitle

\blfootnote{\affilfont$^\textsf{1}$EMBL Barcelona, C/ Dr. Aiguader 88, 08003 Barcelona, Spain.
$^\textsf{2}$EMBL Heidelberg, Developmental Biology Unit, 69117 Heidelberg, Germany.
{\bfseries\sffamily *Author for correspondence: trivedi@embl.es}
}

\vspace{4pt}
\bigskip

%\runningpagewiselinenumbers
%\linenumbers
%\lipsum[1-20]

\textbf{1. Introduction}
\\
Metazoans display vast morphological diversity, yet body plans can universally be distilled to the presence of one to three body axes. Contrasted with protists, a characteristic feature is the existence of a species-specific embryonic phase in the respective life cycle. During its early stages, often collectively termed as gastrulation, simultaneous rearrangement and differentiation transforms a collection of embryonic (stem) cells into a complex, multilayered structure consisting of two (diploblasts) or three (triploblasts) germ layers organized along at least one primary body axis.\smallskip

Despite the conservation of gene regulatory networks that influence cell state and behavior, embryo geometry and overall tissue rearrangement dynamics can vary substantially during axial emergence. Nevertheless, the morphological outcome of this event - a multilayered body plan with its respective axes - remains constant throughout metazoans\green{\cite{keller2003,leptin2005,solnica2005,solnica-krezel2012,mongera2019}}.\smallskip

Traditionally, evolution of the primary axis has been studied from a molecular perspective, by examining the roles of, for instance, Wnt/$\beta$-catenin signaling and T-box genes, which evolutionarily predate the cnidarian-bilaterian split\green{\cite{sebe-pedros2013,holstein2012,loh2016}}. Here, we however suggest that, through examination and comparison of embryo-like \textit{in vitro} and \textit{ex vivo} systems with related entities and native embryos, one may gain deeper insight into the emergence of metazoan body axes as well as the evolution of body plans.\smallskip

Upon consideration of such systems from various species, including \textit{Nematostella}, \textit{Xenopus}, zebrafish, mouse and human, it transpires that isolated ensembles from embryonic stem cells (ESCs) and ESC-like populations (ESC-LPs) universally harbor the capacity to self-organize into a rudimentary body plan with at least a primary axis. Notably, this may occur via developmental trajectories alternative to those in the respective native embryo. We hence argue that these observations could point towards a deeper, evolutionarily \textit{conserved developmental mode} which cells exhibit as they are released from their species-specific geometrical arrangements and mechanochemical signaling environments\green{\cite{newman2016}}.\smallskip

In this perspective, we outline metazoan body axes and conserved initial patterning genes Wnt and Bra/T, followed by a brief review and comparison of mostly recent embryo-like systems in an evolutionary context. Lastly, we speculate on the underlying mechanisms which may have led to the observed diversity in early embryo morphology and on shared, fundamental principles of initial body plan establishment across the animal kingdom.\bigskip

\textbf{2. Animal body axes}
\\
Animals display a variety of body plans consisting of one or more anatomical axes that delineate body polarity and characterize the degree of symmetry in the arrangement of body parts around the axes. Notably, while the initial axis of the pre-gastrulating embryo is often labelled as animal-vegetal, based on the polarity of the oocyte, this may not necessarily correspond to the alignment of the body axes that generally emerge concomitantly with gastrulation or germ layer specification\green{\cite{valentine2004,willmore2012}}.\smallskip

In cnidarians and ctenophores an oral-aboral (OA) axis is ubiquitously identifiable at least in larval stages. This further arguably includes placozoans, demarcated along their respective upper and lower epithelia, the latter of which is used for feeding and exhibits expression of oral-associated genes\green{\cite{dubuc2019}}. Cnidarian polyps and medusae generally exhibit (bi-)radial or bilateral symmetry as opposed to the rotational symmetry of ctenophores\green{\cite{dunn2015,ball2004}}. Larval sponges feature an antero-posterior (AP) axis that is alternatively labelled animal-vegetal (AV) axis as well as radial symmetry. In contrast, adult sponges only display an apical-basal (AB) axis, due to clear absence of a morphological mouth and indistinct body symmetry\green{\cite{swalla2006,degnan2005,ryan2007, adamska2007,petersen2009,genikhovich2017}}.\smallskip

%%%%%%%%%%%%%%%%%%%%%%%%%%%%%%
%%%%%%%%%%%%%%%%%%%%%%%%%%%%%%%
\begin{figure*}[ht]
\centering
\includegraphics[width=7in]{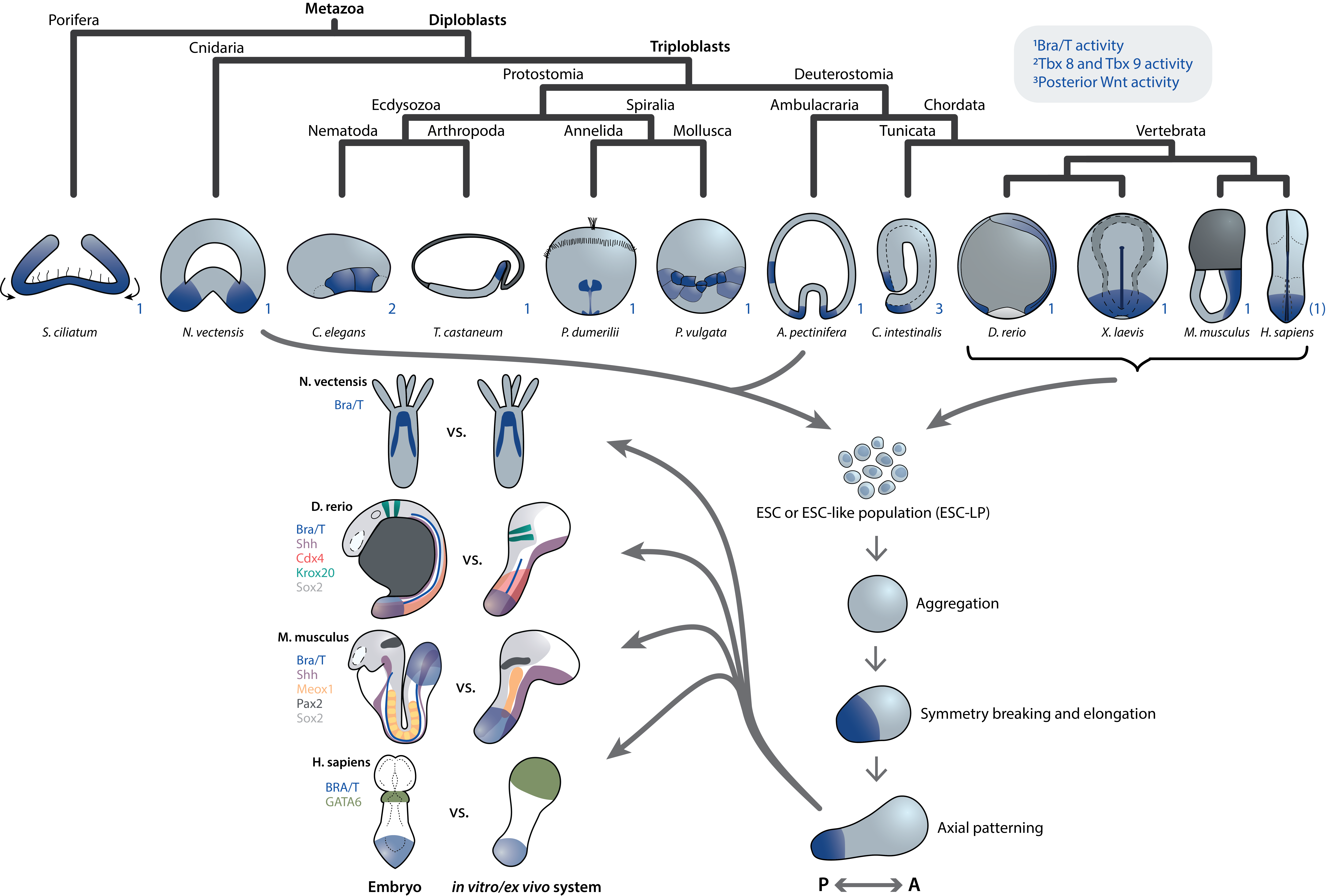}
 \caption{{\fignumfont Primary axis formation during gastrulation in metazoan embryos and corresponding artificial systems.} Metazoan embryos around the gastrulation or an equivalent developmental phase exhibit distinct morphologies and associated overall tissue rearrangement dynamics. Yet, during this event they universally specify at least a primary body axis, demarcated by conserved expression of posterior (or oral in cnidarians) patterning determinants. Among them are T-box and Wnt genes, localized transcriptional activity of which is highlighted in blue. Note that marker gene expression patterns in the human embryo are speculative.
 Research efforts have shown that ESC or ESC-like populations from species across the animal kingdom can be re-aggregated \textit{in vitro} and, although lacking the respective external environment and associated developmental cues, remain capable of recapitulating at least a basic transcriptional body plan with an anteroposterior (AP) and oral-aboral (OA) axis, respectively.
 Strikingly, comparison of examples for such in vivo or ex vivo systems highlights a remarkable overall similarity despite the varying geometry of the respective native embryo. This may point towards the existence of a conserved developmental mode that cells exhibit when released from their species-specific extraembryonic environment.
 \label{Fig1}}
\end{figure*}
%%%%%%%%%%%%%%%%%%%%%%%%%%%%%%%
%%%%%%%%%%%%%%%%%%%%%%%%%%%%%%%

As the second major group next to the diploblasts, all other extant animals are referred to as triploblasts or bilaterians. While there is considerable variability, the body plan of most bilaterian phyla features three distinct axes: Anteroposterior (AP), dorsoventral (DV) and mediolateral (ML, also labelled as the midline) as well as bilateral symmetry\green{\cite{genikhovich2017}}.\smallskip 

Moreover, as previously alluded to, although adult body plans can be congruent with the axial emergence during earlier embryonic phases, this is not generalizable across metazoans: While the early gastrula of \textit{Nematostella} exhibits radial symmetry around the OA axis, bilateral symmetry takes over at late gastrula stages\green{\cite{genikhovich2017}}. This is further exemplified by echinoderms, where, in contrast to their larval form, adult starfish are radially symmetric with an OA axis\green{\cite{ji2012}}.\smallskip

Beyond the natural anatomical meaning and the morphological significance (feeding and digestion), studies have highlighted the convergent regulatory logic of AP and OA axis establishment via a Wnt/$\beta$-catenin-dependent system in deuterostomes and cnidarians\green{\cite{bagaeva2020}}. Despite the scarcity of experimental data in case of ctenophores, placozoans and sponges, it is still reasonable to assume hat the AP axis of bilaterians, the OA axis of cnidarians, ctenophores and placozoans as well as the AP or AV axis of sponges constitute the most evolutionary ancient and thus main or primary body axis in these organisms.\smallskip 

Apart from Wnt/$\beta$-catenin, transcription factor Bra/T is another well-studied marker of primary axis, and thus initial body plan patterning\green{\cite{petersen2009,swalla2006}}. To illustrate their deep evolutionary conservation, we will briefly elaborate on the role of these factors in organisms from different phyla and speculate on resulting implications for the evolution of initial body plan patterning.

%%%%%%%%%%%
\medskip

\textbf{2.1 Wnts in primary axis specification}
\\
Wnts are a family of secreted glycoprotein ligands that, via binding to Frizzled transmembrane receptor proteins, activate eponymous pathways implicated not only in early embryonic axial patterning but also a plethora of contexts, including general regulation of cell fate specification and proliferation. Intriguingly, Wnt genes appear to be exclusive to metazoans, having not been identified in unicellular eukaryotes, plants or fungi\green{\cite{croce2008,holstein2012}}. Hence, the emergence of Wnt has been intricately linked to the emergence of animal multicellularity due to its implication as a facilitator of symmetry breaking on the single-cell as well as on the tissue level\green{\cite{loh2016}}.\smallskip

Historically, Wnt signaling has been divided into canonical and non-canonical Wnt signaling. Central to the former is the $\beta$-catenin which is stabilized via inhibition of key regulatory kinase GSK-3B as part of a signaling cascade elicited through binding of Wnt to Frizzled. $\beta$-catenin is a multifunctional protein that is also directly involved in cell adhesion by being part of the complex which links cadherins to the actin cytoskeleton. Moreover, in contrast to Wnt genes themselves, both GSK-3$\beta$ and $\beta$-catenin homologs have been identified in all eukaryotes.

\begin{figure*}[ht]
\centering
\includegraphics[width=7in]{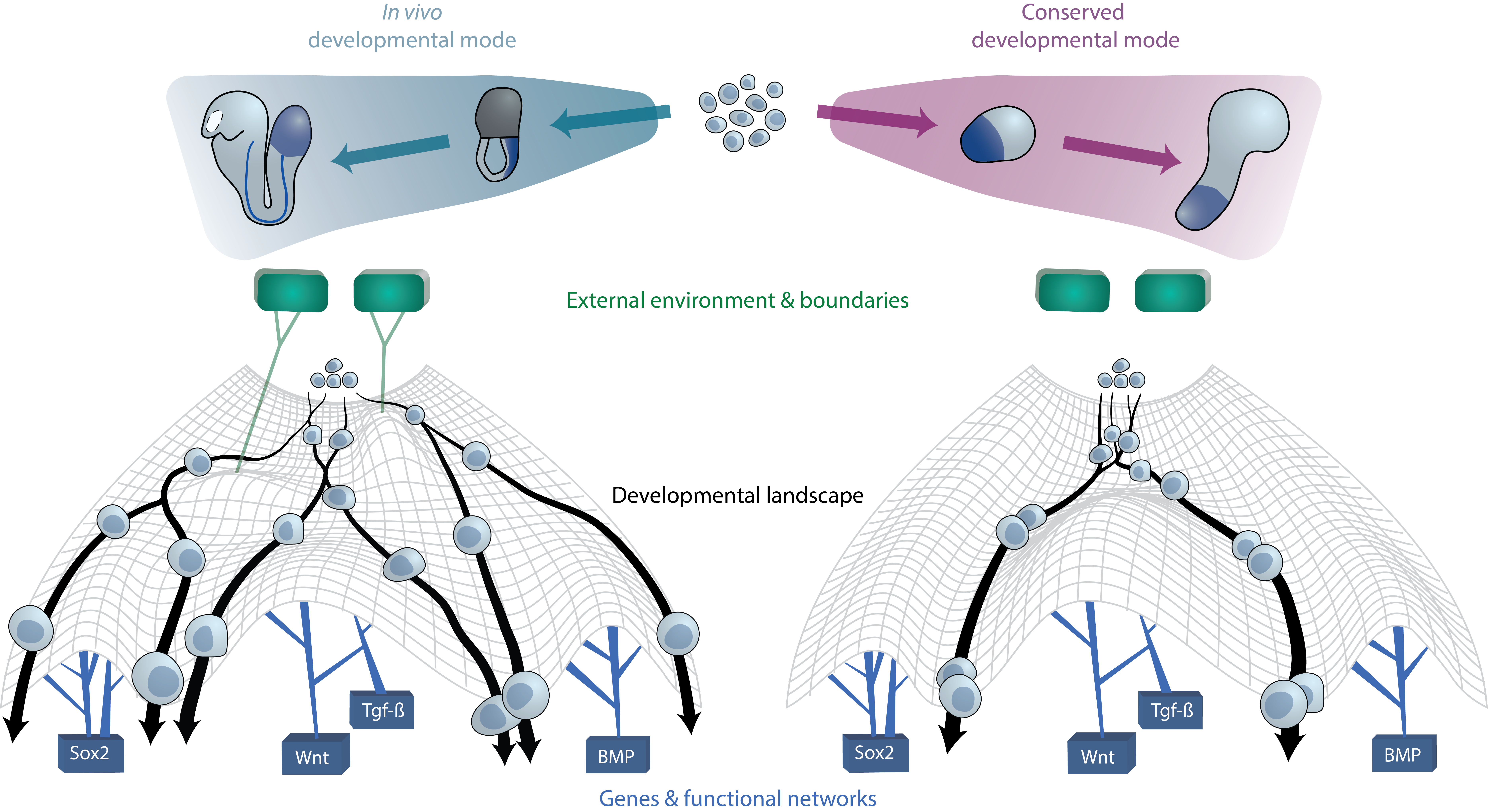}
 \caption{{\fignumfont Diverse developmental trajectories accessible by ESCs and ESC-LPs in the presence and absence of extraembryonic inputs.} The developmental trajectories which aggregates of ESC-LPs exhibit upon removal of external or extraembryonic and associated boundaries may constitute a conserved mode that is shared across species. On a cellular level, this can be visually approximated as cells undergoing differentiation within Waddington's developmental landscape\green{\cite{waddington1957}}. The landscape is shaped by key gene networks which remain constant between species and \textit{in vitro} (bottom) as well as the external (micro-)environment and embryo geometry (top), here represented as green tiles, which vary between species. In case the latter factors are not present as ESCs are removed from their native context and grown \textit{in vitro}, cellular developmental trajectories revert to the aforementioned conserved mode since cells from different species now experience the same landscape.
 \label{Fig3}}
\end{figure*}

\begin{tcolorbox}[title=\textbf{Box definitions and abbreviations 1}]
	\textbf{Metazoans (Animals)}
	\\
	Multicellular eukaryotes which are heterotrophic, i.e.\ they cannot produce their own food and thus depend on external organic material for nourishment.\ They are motile during at least part of their life-cycle.
	\medskip
	
	\textbf{Phylum (plural: Phyla)}
	\\
	A rank in the taxonomic hierarchy used to classify similar groups of biological organisms. Phylum is the rank below kingdom (e.g.\ metazoans, protists, plants, fungi) and above class. For instance, humans belong to the kingdom \textit{metazoa}, the phylum \textit{chordata} and the class \textit{mammalia}.
	\medskip
	
	\textbf{Body plan}
	\\
	A set of morphological features describing the body shape and structure of a given species. The most fundamental aspects of body plans, such as body axes, are conserved even across phyla.
	\medskip	
	
	\textbf{ESCs and ESC-LPs}
	\\
	Embryonic stem cells and embryonic stem cell-like populations.
	\medskip	
	
	\textbf{Self-organization}
	\\
	Emergence of order  based on local interactions between different parts of a system that was initially disordered.\ Characteristic of such a process is feedback between components that amplifies the effects of local interactions or disturbance to the global level.\ In the context of aggregation experiments of ESCs and ESC-LPs discussed in this review, emergence of the primary axis, can be considered self-organization.\ A related concept in biological systems is that of \textit{genetically encoded self-assembly}\green{\cite{turner2016}}, meant to imply processes where the information for the emergence of order is already encoded in the constituent cells that are primed to form structures or patterns under the influence of genetic programs\green{\cite{xavier2019}}.
	\medskip
	
	\textbf{Symmetry breaking}
	\\
	Emergence of an asymmetry, morphological or molecular, i.e.\ in terms of gene expression, within a previously homogenous structure.\ Here, we refer to symmetry breaking in the context of asymmetries at the level of cell populations rather than asymmetries displayed by single cells of an aggregate.\medskip
	
	\textbf{Conserved developmental mode}
	\\
	Cell ensembles from ESCs grown in absence of species-specific external (micro-)environments and cues adopt a set of differentiation trajectories in terms of both morphogenesis and gene expression which are evolutionarily conserved across different species, in order to form patterned structures resembling actual embryos (Fig.\ \ref{Fig3}).\ A similar term, \textit{dynamical patterning} or \textit{morphogenetic module} (DPMs), refers to the conserved set of molecules which are the products of those genes, as well as their physical effects, that alter the state, shape, size and arrangement of cells in a given population\green{\cite{newman2016}}.\ Therefore, several of such DPMs can be activated as ESCs or ESC-LPs revert to the conserved developmental mode.
\end{tcolorbox}

It is noteworthy that mechanical strains developed by tissue movement during gastrulation in bilaterians have been shown to trigger the phosphorylation of $\beta$-catenin, thereby imparting early mesodermal identity to the cells in an evolutionary conserved manner. Such a mechanosensitive translocation of $\beta$-catenin away from the membrane induces expression Bra/T orthologue tbxta in zebrafish and Twist in \textit{Drosophila} thereby generating mesoderm in the gastrulating embryo\green{\cite{brunet2013}}.\smallskip

In contrast, non-canonical Wnt signaling traditionally comprises $\beta$-catenin independent Wnt pathways. These mainly include the planar cell polarity (PCP), regulating cellular migration and cytoskeletal asymmetry, as well as the Calcium-mediated pathway which acts via release of $Ca^{2+}$ and subsequent activation of Ca-dependent enzymes. Amongst others, these have been found to be involved in various developmental events, including axial extension and organ formation\green{\cite{de2011}}.\smallskip

Whereas it was previously assumed that the 3 main Wnt pathways operate separately, research from the last two decades favours an integrated view of Wnt signaling, given that its individual components can be involved in several contexts\green{\cite{komiya2008,nayak2016}}.\smallskip

Throughout metazoans, a subset of Wnt activity thus marks the posterior in bilaterian embryos and sponge larvae or the oral pole in ctenophores and cnidarians\green{\cite{cadigan1997,leininger2014,petersen2009, yamada2007,yamada2010,mcgregor2008,hogvall2019}}.\ Adult sponges exhibit Wnt expression in apical tissues which are thought to be generated by the larval posterior pole\green{\cite{windsor-reid2018}}. While genomic analysis of the placozoan \textit{Trichoplax adhaerens} has revealed components of canonical Wnt signaling (Wnt genes, Dsh, Frz, GSK3, Axin, $\beta$-catenin TCF), the spatial expression patterns of Wnt genes and $\beta$-catenin to date remain unknown\green{\cite{srivastava2008}}.\smallskip 

Albeit Wnts are among the incipient primary axis specifiers, the origin of embryonic axial polarity in many species, including \textit{C.\ elegans}, insects, ascidians and anamniotes appears to stem from localized maternal deposition of fate determinants\green{\cite{yan2018,prodon2005,heasman2006,king2005,sardet2003,rose2014}}. The \textit{Drosophila} egg constitutes a particularly well-studied example in which, prior to fertilization, maternal effect genes bicoid and nanos become localized to the anterior and posterior tip, respectively, thereby constituting the initial AP polarization event\green{\cite{kimelman2011}}.\smallskip 

While \textit{Drosophila} Wnt1 ortholog wingless (Wg) fulfils later roles in segment polarity establishment, midgut morphogenesis and limb development, its expression surfaces already before gastrulation, during early cellularization of the blastoderm, initially in a the shape of a band at the posterior of the embryo\green{\cite{vorwald-denholtz2011,swarup2012}}.\smallskip

Unlike \textit{Drosophila}, in studied long germ band insects, including \textit{Tribolium castaneum}, Wnt expression at the posterior is maintained in some cells of the posterior growth zone during the characteristic secondary segmentation process. Hence, \textit{Drosophila} and other short germ band insects likely lost this Wnt domain as they specify all their segments almost simultaneously\green{\cite{bolognesi2008_1,bolognesi2008_2}}.
%%%%%%%%%%%%%%%%%%%%%%%%%%%%%%%
%%%%%%%%%%%%%%%%%%%%%%%%%%%%%%%
\bigskip

\textbf{2.2 T-box genes and Brachyury}
\\
Brachyury or T (Bra/T) is the most ancient, founding member of the T-box transcription factor family, characterized by their eponymous, DNA-binding N-terminal T domain of about 180-200 amino acids\green{\cite{showell2004,papaioannou2014}}. During gastrulation, Bra/T defines and patterns the blastopore in cnidarians and ctenophores, giving rise to the oral region, as well as the posterior and incipient mes(endo-)doderm in triploblasts\green{\cite{swalla2006,yamada2007}}. In early S. ciliatum sponge embryos, elevated Bra/T transcriptional activity is localized to a region undergoing a characteristic inversion event which shapes the thus emerging larva\green{\cite{leininger2014}}.\smallskip

Collectively, data in sponges and diploblasts argue that an ancient role of Bra/T is to regulate the morphogenetic movements such as invagination or folding during gastrulation\green{\cite{technau2001,yamada2010}}.\smallskip 

Indeed, Bra/T is expressed at the site of gastrulation in the vast majority of studied species, spanning a plethora of (sub-) phyla, including ctenophores, cnidarians, xenacoelomorpha, molluscs, chaetognatha, arthropods, ambulacraria and chordates\green{\cite{scholz2003,lartillot2002,takada2002,hejnol2008,arenas-mena2013,tagawa1998,shoguchi1999,holland1995,kispert1994,beddington1992,marcellini2003,smith1991,kispert1995,arendt2001,satoh2000}}, in the latter of which it is further crucial for notochord specification and development\green{\cite{papaioannou1998}}. Intriguingly, \textit{C.\ elegans}, a nematode, does not have a clearly identifiable Bra/T ortholog, despite featuring an impressive array of 20 predicted T-box genes that are comparatively diverged. Yet, during the initial stages of \textit{C.\ elegans} gastrulation, expression of T-box family members Tbx8 and Tbx9 is restricted to two endodermal precursor cells which are the first to internalize\green{\cite{pocock2004}}. At later stages, mab9, another T-box gene related to Brachyury is required for posterior hindgut formation\green{\cite{woollard2000}}.\smallskip

In insects, Bra/T ortholog Brachyenteron is required for the same process and further for specification of caudal visceral mesoderm\green{\cite{kispert1994,kusch1999,singer1996,shinmyo2006}}. Earlier in embryogenesis, during the blastoderm stage, Bra/T expression surfaces in the posterior part of the embryo. At the onset of gastrulation, it can be identified in the posteriormost region of the presumptive mesoderm that is subsequently internalized. While in \textit{Drosophila} Bra/T expression diminishes in these cells, it is maintained in the long germ band insect \textit{Tribolium} throughout the posterior growth and segmentation phase of its extending germ band, similar to posterior expression dynamics of Wnt in the two species\green{\cite{berns2008}}. Likewise, given that \textit{Drosophila} as a short germ band insect lacks a secondary growth phase due to specifying almost all segments simultaneously, it seems probable that Bra/T activity was superfluous in this context and therefore lost.\smallskip

Emergence of body axes is concomitant with the specification of the germ layers and therefore the role and localization of T has also evolved in this context. \textit{Hydra} features two Bra paralogues, Hybra1 and 2 which exhibit distinct expression patterns during head formation, the former being predominantly localized to the endoderm and the latter to the ectoderm\green{\cite{technau1999,bielen2007}}. In studied cnidarian embryos, including \textit{Nematostella} and \textit{Acropora}, Bra/T is expressed around the blastopore during gastrulation, defining the boundary between ecto- and endoderm and is required for patterning thereof\green{\cite{scholz2003,yasuoka2016}}. With the emergence of triploblastic animals, Bra/T became a universal marker for incipient mesodermal identity during early development.\smallskip 

Intriguingly, while endogenously endodermal Hybra1 has been shown to induce mesoderm in the vertebrate \textit{Xenopus}, endogenously ectodermal Hybra2 promotes development of neural tissues\green{\cite{bielen2007}}. An analogy can be drawn to neuromesodermal progenitors (NMPs) of amniote embryos, where Bra/T$ ^{+} $ cells also give rise to ectodermal tissue. NMPs are bi-potent stem cells co-expressing Bra/T and Sox2, which contribute to spinal cord tissue and pre-somitic mesoderm (PSM) during axial elongation \green{\cite{henrique2015,steventon2017}}.\smallskip

Taken together, Bra/T is a gene at the interface of ecto- meso- and endoderm: While generally associated with mesoderm formation in bilaterians, depending on the context and the species studied, it may be involved in specification of any of the three germ layers. What remains consistent, however, is the implication of Bra/T in cell movements or adhesion properties\green{\cite{burton2008,technau2003}}.

The emergence of animal multicellularity has been linked to Wnt, as previously mentioned, and TGF-$\beta$ signalling. Notably, while in several studied species Bra/T has been found to act immediately downstream of these two pathways\green{\cite{arnold2000,vonica2002,turner2014,latinkic1997,pauklin2015,cerdan2012}}, T-box transcription factors themselves predate multicellularity, having been identified in several unicellular opisthokonts, including several fungi taxa as well as ichthyosporeans and filastereans\green{\cite{sebe-pedros2013}}.\smallskip

Strikingly, binding specificity of these Bra/T homologues is highly conserved, as functionally exemplified by injection of Bra/T from \textit{Capsaspora owczarzaki}, an amoeiboid filasterean, rescuing gastrulation-defect phenotypes in Bra/T-deficient \textit{Xenopus} embryos. On the other hand, nonmetazoan Bra/Ts do not appear to undergo cofactor interactions of metazoan Bra/Ts, thereby suggesting that extensive subfunctionalization of T box genes at the origin of metazoans was achieved via co-option of external genes into regulatory networks.\smallskip

Moreover, in contrast to sponge T-box homologues, Bra/T from \textit{C. owczarzaki} is able to induce endodermal lineage determinants Sox17 and endodermin in \textit{Xenopus}. This suggests that sponge Bra/T has indeed undergone subfunctionalization and further implies that the mesoderm might have emerged through integration of Bra/T with other -novel or existing- genes into mesoderm-specifying modules\green{\cite{sebe-pedros2013}}.

%%%%%%%%%%%%%%%%%%%%%%%%%%%%%%%
%%%%%%%%%%%%%%%%%%%%%%%%%%%%%%%
\bigskip

\textbf{3. Evolution of body plans \textit{in vivo} and the significance of systems from ESCs and ESC-LPs}
\\
Despite the evolutionarily conserved role of Wnt and T/Bra in patterning of the primary body axis, the diversity of shape, size, dynamics of tissue rearrangements and the environmental niche across embryos of different species\green{\cite{leptin2005,solnica-krezel2012}} points towards a vast space of functional developmental trajectories accessible to early embryonic cells. Furthermore, both within and between phyla there is evident topology shuffling within key developmental gene networks, especially in terms of co-option of novel or external genes into such regulatory modules, thereby creating novel developmental mechanisms via the cell-to-tissue-level morphogenetic processes they influence\green{\cite{chen2013,zhang2015}}.\smallskip 

Such a perspective hints towards a picture of evolution that transcends the simple conservation of genes to the conservation of dynamic developmental modes which unfold to generate similar axial coordinates across phyla, while the constituent cells themselves might take diverse developmental trajectories dictated by the embryonic and extraembryonic environment. This notion is also encapsulated in the established concept of \textit{dynamical patterning modules} (DPMs), referring to evolutionary conserved sets of gene products as well as their physical effects that alter the state, shape, size and arrangement of cells in a given population\green{\cite{newman2016}}.\smallskip

Understanding these developmental modes \textit{in vivo} comes with the challenge that cells within the embryonic context are also constrained by distinct factors - embryonic (shape) and extraembryonic (signaling and mechanical cues) - that generate the observed morphological diversity. We note that such ``constraints'' do not imply less sophisticated developmental processes, on the contrary, in many species, external input is strictly required for complex morphogenesis to occur\green{\cite{sheng2012,christodoulou2019}}.\smallskip

Therefore the study of \textit{ex vivo} or \textit{in vitro} systems can elucidate, on the one hand, shared principles and, on the other hand, provide an alternative angle: Removing stem cells (SCs) and SC-like populations from their native context potentially allows the disentanglement of such species-specific factors, thereby allowing cells to follow developmental trajectories guided by their inherent self-organizing capabilities \green{\cite{sogabe2019}}.\smallskip

This concept is well exemplified by the various kinds of embryo-like structures, termed embryoids or embryonic organoids that have been developed recently from either embryonic or induced pluripotent stem cells (iPSCs) to recapitulate patterning events of initial embryogenesis until the onset of organogenesis\green{\cite{shahbazi2019,baillie-benson2020,moris2020_2}}.\smallskip

Historically and perhaps conceptually, they are preceded by SC embryonic explant and re-aggregation systems\green{\cite{wilson1907}}, several of which have been established already decades ago in evolutionarily distant species such as \textit{Hydra} and starfish\green{\cite{wilson1911,gierer1972,dan-sohkawa1986}}. These and related research efforts have underscored the regenerative capacity as well as the self-organizing potential of ESC-LPs across metazoans, given that, in both species, and thus across clades, re-aggregated cells are able to reconstitute a functional animal. Concomitantly with the emergence of new \textit{in vitro} embryo-like structures, a few novel examples of such \textit{ex vivo} systems have also been reported as of late.\smallskip

In what follows, we will list mostly recent examples from both diploblasts and triploblasts and discuss selected ones as to how far these can provide clues for elucidating universal principles of animal body axis specification. While we aim to present an evolutionary perspective with as much phylum coverage as possible, it has to be noted that relevant data on non-mammalian systems is rare and limited to re-aggregation approaches.

\bigskip
\textbf{3.1 Re-aggregation Studies in cnidarians}
\\
In the cnidarian model organism \textit{Hydra}, single cell dissociation and reaggregation experiments were already performed decades ago and yielded insights into regeneration and tissue developmental processes\green{\cite{wilson1911,gierer1972}}. Remarkably, \textit{Hydra} exhibits vast regenerative potential with any extracted fragment of its body being able to reconstitute an animal\green{\cite{vogg2019}}. Labelling experiments have demonstrated that pattern formation in re-aggregates arises de novo and that cells sort themselves only in regards to their germ layer identity\green{\cite{technau1992,technau2000}}.\smallskip 

Additional molecular characterization has revealed that around 24h post-aggregation, HyWnt-expressing spots appear in the \textit{Hydra} re-aggregates which later co-localize with head protrusions where Hy-$\beta$-Cat and HyTcf are transcriptionally upregulated\green{\cite{hobmayer2000}}. Such a self-organizing activity is implied to result from autoregulation and repression of HyWnt3\green{\cite{nakamura2011}} that is coupled to mechanical stress\green{\cite{futterer2003}} and together triggers the symmetry-breaking\green{\cite{soriano2009,mercker2015}} event crucial for establishing a Wnt-expressing organizing center.\smallskip

A more recent study analyzes the dissociation and re-aggregation of \textit{Nematostella} mid-gastrulae in order to elucidate developmental plasticity of embryonic cell populations\green{\cite{kirillova2018}}. Oral halves, which contain a population of blastopore lip cells with axial organizer capability conveyed by Wnt1 and Wnt3, were able to reorganize the body plan and develop into a functional polyp. Accordingly, aboral halves were only capable of reconstituting an animal upon prior injection of both Wnts, demonstrating that a competent cell population (ESC-LPs) is needed for generation of a re-aggregate capable of developing into a functional animal.\smallskip 

Furthermore, these experiments suggest that this competence can be retroactively conveyed by addition of relevant factors into cells which have previously undergone a separate differentiation trajectory, similar to the reprogramming strategies used for induced pluripotent stem cells\green{\cite{takahashi2006}}. Therefore, a stem cell population in a more naïve state, from embryos prior to the gastrula stage, should, in theory, harbour the full developmental potential to self-organize the body axes without external inputs.\smallskip

Akin to corresponding experiments in adult \textit{Hydra}, cells in developing embryonic \textit{Nematostella} re-aggregates exhibit vast reprogramming of their axial identity. Notably, while ectodermal cells could convert into endoderm, the latter remained endodermal. However, in case re-aggregates were generated purely from endoderm, cells became mesenchymal and migratory causing aggregates to collapse.\smallskip

Perhaps most intriguing though is the observation that re-aggregates which successfully reform a functional animal employ a developmental mode distinct to normal \textit{Nematostella} embryogenesis, but similar to other cnidarians, such as hydrozoans. Namely, instead of invagination, germ layer specification occurs via delamination of the ectoderm, ingression of those endodermal plate cells that were initially located at the aggregate surface due to random mixing, as well as cavitation of inner cells.

%%%%%%%%%%%%%%%%%%%%%%%%%%%%%%%
%%%%%%%%%%%%%%%%%%%%%%%%%%%%%%%
\medskip

\textbf{3.2 Explants in Amniotes}
\\
The field of amphibian and avian development has had a long history of excising tissues from the native embryo and either transplanting them in a different embryo or different regions of the same embryo or culturing them as \textit{ex vivo} explants over extended periods\green{\cite{spemann1923,mangold1933,holtfreter1933,waddington1932,mookerjee1953,ball1966,hamburger1969,gilbert1991}}. While these studies have provided crucial insights into the developmental potential of cells through the comparison of \textit{ex vivo} explants from different stages of development\green{\cite{keller1988}}, a detailed picture of the self-organizing capabilities of cells, akin to reaggregation studies in cnidarians or 3D aggregates of ESCs from mammals, is emerging only recently.\smallskip

In \textit{Xenopus}, cell aggregates obtained by dissociating chordamesoderm (prospective notochord) from the early gastrula were shown to undergo cell sorting and convergent extension to establish an anteroposterior patterning in an elongated structure revealed through the expression of Bra/T (Xbra) and Chordin\green{\cite{ninomiya2004}}. A more drastic example is that of aggregates of cells from dissociated animal cap explants, at an earlier developmental stage. By addition of Activin A to these cells, they were shown to generate multiple mesodermal tissue types\green{\cite{green1990}} and, depending on the dosage of the signal, they consist almost entirely of notochord\green{\cite{green1992}}. Perhaps even more remarkable is the scenario when such dissociated cells are allowed to re-aggregate spontaneously: They round up into a ball followed by elongation and organization of an axis in a manner whereby the axial extrusion resembles archenteron elongation in deuterostomes such as sea urchins \green{\cite{green2004}}.\smallskip

Similar to amphibians, different types of explants from teleost embryos have also been used historically to explore the self-organization of cells. It was shown that isolated blastoderms of \textit{Fundulus heteroclitus}, deprived of yolk, can develop into embryo-like structures \green{\cite{oppenheimer1936}}. Furthermore, uncommitted embryonic cells of animal caps in 128 cell stage zebrafish embryos were shown to organize a complete embryonic axis upon injection of opposing gradients of BMP and Nodal\green{\cite{xu2014}}.\smallskip 

Recently an \textit{ex vivo} system - termed pescoids\green{\cite{trivedi2019}} in analogy to mammalian gastruloids, utilized most of the blastodermal cells severed from yolk in zebrafish embryos around 256 cells stage, before germ layer induction. These \textit{ex vivo} systems when allowed to grow without the supply of external signals were shown to be robust to cell mixing, able to form a polarized domain of high Bra/T (tbxta) expression, leading to further elongation under the influence of Nodal (similar to amphibian requirement for Activin) and non-canonical Wnt (PCP) signalling. In addition, they specify all germ layers and most mesodermal lineage precursors in terms of marker gene expression\green{\cite{fulton2020,schauer2020,williams2020}}.\smallskip 

Results in amniotic embryos need to be interpreted with care since the embryonic cells inherit maternal signals starting at the single cell stage, as it was indeed shown for pescoids that require polarized inheritance of maternal factors\green{\cite{schauer2020}}. A definite demonstration of spontaneous self-organization requires availability of ESC-LPs for teleosts and efforts towards culturing ESCs for medaka will prove to be a valuable tool in this regard\green{\cite{hong1996a,hong1996b,hong1998,hong2011}}.\smallskip 

In their natural settings, the phenomenon of complete dispersion of pre‐embryonic blastomeres in annual killifish (\textit{Austrofundulus myersi}\green{\cite{wourms1972}}, \textit{Austrolebias charrua}\green{\cite{pereiro2017}}), followed by their re-aggregation to generate embryonic body plan is the closest observation to complete dissociation of teleost cells. While it is not clear to what extent the organization of dissociated cells is autonomous and independent of extraembryomic cues, annual killifish, intriguingly, represent an exceptional \textit{in vivo} testimony to the notion of multiple developmental trajectories being available to the embryonic cells as seen in the differential rate of specification of anterior and posterior structures in embryos that have and have not undergone diapause\green{\cite{podrabsky2010,romney2018}}. 

\begin{tcolorbox}[title=\textbf{Box definitions and abbreviations 2}]
	\textbf{TSCs}
	\\
	Trophoblast stem cells. They represent the first extra-embryonic lineage to be specified in mammals, crucial for implantation into the uterine wall and placenta formation.
	\medskip
	
	\textbf{XEN cells}
	\\
	Extraembryonic endoderm cells. Another extraembryonic lineage in mammals, derived from the hypoblast, the second extraembryonic lineage, which is formed as the inner cell mass (ICM) of the blastocyst, surrounded by the trophoblast, differentiates into the epiblast, giving rise to the embryo proper and the hypoblast. In mouse, a subpopulation of extraembryonic endoderm has been shown to be critical for AP axis formation.
	\medskip
	
	\textbf{EPSCs}
	\\
	Extended pluripotent stem cells. They can give rise to both extraembryonic and embryonic lineages, as opposed to ESCs which are limited to the latter.
	\medskip
	
	\textbf{Blastoids}
	\\
	An \textit{in vitro} model or embryo-like structure made from mESCs and mTSCs resembling the mouse blastocyst, the native embryo around stage E3.5.
	\medskip
	
	\textbf{ETS and ETX embryos}
	\\
	\textit{In vitro} models or embryo-like structures made from mTSCs and mESCs (ETS embryo) and mTSCs, mESCs as well as mXEN-cells (ETX embryo), respectively. They closely mimic mouse embryo geometry and morphogenesis until around E7.0-7.5.
	\vspace*{0.2cm}
	
	\textbf{Gastruloids}
	\\
	An \textit{in vitro} model or embryo-like structure made from mESCs (mGastruloids) or hESCs (hGastruloids) in low-adherence, minimal conditions, recapitulating aspects of embryogenesis, in particular the development of spatially distinct gene expression domains demarcating the basic body plan up to E9.0-9.5 in case of mGastruloids, while apparently lacking complex structures and morphogenesis without external input. 
	\vspace*{0.2cm}
	
	\textbf{MDT}
	\\
	Mid-developmental transition.\ A phase in mid-embryogenesis during which species from the same phylum exhibit convergent gene expression profiles, distinguishing themselves from other phyla.
\end{tcolorbox}

%%%%%%%%%%%%%%%%%%%%%%%%%%%%%%%
%%%%%%%%%%%%%%%%%%%%%%%%%%%%%%%
%\bigskip

\textbf{3.3 Mammalian \textit{in vitro} systems}\vspace*{0.1cm}
\\
In the past few years, a diverse array of mammalian \textit{in vitro} embryonic models has emerged, generated from either mouse or human ESCs\green{\cite{shahbazi2019,baillie-benson2020}}. Pioneering work on 3D aggregates, termed embryoid bodies (EBs), has demonstrated that these can give rise to progenitor cells for the germ layers as well as form rudiments of tissues and organs without the context of an embryo\green{\cite{doetschman1985,desbaillets2000,hoepfl2004,gadue2006,kubo2004,brickman2016}}. Polarized gene expression and self-organized axial emergence in embryoid bodies has been shown to be mediated by Wnt signalling\green{\cite{tenberge2008,sagy2019}}, although the extent of displayed axial organization is limited.\medskip

\textbf{Mouse gastruloids}
\\
mGastruloids are initially spherical aggregates derived from a few hundreds of homogenous mESCs which, although removed from extraembryonic tissues and nearly all associated signalling and mechanical cues, mimic some morphogenetic events of early mouse embryos, such as elongation, germ layer as well as axis formation and associated patterning\green{\cite{vandenbrink2014,moris2020}}. Initially mGastruloids were developed motivated by the observation that, under differentiation conditions, mouse embryo P19 carcinoma cells are able to form polarized, elongated structures\green{\cite{marikawa2009}}. Recently, they have also been extrapolated to hESCs (hGastruloids)\green{\cite{moris2020}}.\smallskip

AP axis specification in aggregates is demarcated by autonomous polarization of previously homogeneously distributed canonical Wnt and Bra/T expression, constituting the first system-wide symmetry breaking event identified thus far\green{\cite{turner2017}}. After 6-7 days in non-adherent culture, mGastruloids not only develop transcriptionally demarcated AP, DV and ML axes, but also display spatiotemporal expression of hox gene clusters similar to those in mouse at roughly embryonic day (E) 9.0-9.5\green{\cite{beccari2018}}.\medskip

%AP axis specification in aggregates is demarcated by autonomous polarization of previously homogeneously distributed canonical Wnt and Bra/T expression, constituting the first system-wide symmetry breaking event identified thus far.\ As to the endoderm, recent studies have shown that cells expressing markers such as Sox17 and Cdh1 arise dispersed and subsequently congregate without the requirement of an epithelial-to-mesenchymal transition (EMT), like in the native embryo\green{\cite{vianello2020,hashmi2020}}.\ After 6-7 days in non-adherent culture, mGastruloids not only develop transcriptionally demarcated AP, DV and ML axes, but also display spatiotemporal expression of hox gene clusters similar to those in mouse at roughly embryonic day (E) 9.0-9.5\green{\cite{beccari2018}}.\medskip 

\noindent\textbf{Human gastruloids}
\\
Recently established hGastruloids constitute a model for human early AP patterning\green{\cite{moris2020}}. Morphologically resembling their mouse counterpart, they are generated from few hundreds of hESCs, pre-treated with a WNT agonist for 24h in 2D culture, then aggregated in low adherence, differentiation conditions with ongoing WNT upregulation that is subsequently diluted. BRA/T develops polarized expression a day after aggregation, thereby demarcating the posterior pole which subsequently elongates.\smallskip 

Already after 3-4 days of development, hGastruloids reach maximal elongation and exhibit expression of genes associated with all 3 germ layers as well as a plethora of AP axial patterning genes, spatially organized in a manner similar to mammalian embryos. These notably include anterior cardiac mesoderm- and posterior node-related genes as well as a posterior to anterior somitogenesis signature. Hence, in terms of their transcriptional profile, hGastruloids partly correspond to CS9 human embryos. Akin to mGastruloids, for the most part, they do not seem to recapitulate early human-specific embryonic morphology.\smallskip 

%%%%%%%%%%%%%%%%%%%%%%%%%%%%%%%
%%%%%%%%%%%%%%%%%%%%%%%%%%%%%%%
\bigskip

\textbf{4. Common developmental trajectories accessible by ESC(-LP)s across species} 
\\
Despite the differences in early embryogenesis of source species, studies involving aggregates of corresponding ESC(-LPs) in \textit{Nematostella}, zebrafish, \textit{Xenopus}, mouse and human embryonic model systems, reveal several unifying observations as well as interpretations. Particularly with respect to the establishment of the primary axis, cell aggregates develop an oral or posterior domain characterized by Bra/T expression in the native embryo, demarcating a symmetry-breaking event \red{(Fig.\ \ref{Fig1})}, followed by axial elongation and AP patterning.\smallskip

In addition, a morphological similarity - initially spherical, then elongated - can be observed between the different systems, notwithstanding the strikingly distinct geometries of the \textit{Nematostella}, fish, mouse and human embryos. While the precise mechanism remains to be elucidated, another commonality is the analogous dependence of primary axis formation on the initial number of cells in aggregates from two evolutionarily highly distant species: Larger re-aggregates from \textit{Nematostella} gastrulae\green{\cite{kirillova2018}} as well as mouse ESCs (gastruloids\green{\cite{vandenbrink2014}} and embryoid bodies\green{\cite{tenberge2008}}) form multiple oral and posterior domains, respectively.\smallskip

Altogether, aforementioned developmental similarities seem to reflect an inherent, converging self-organizing capability of ESCs and ESC-LPs across species when developing in the most minimal environment with the least possible external biochemical and mechanical input.\smallskip 

We argue that this capability represents an evolutionarily conserved developmental mode which we define as the set of developmental trajectories - in terms of morphogenesis and gene expression dynamics - that cell collectives within a given system or aggregate from ESCs undergo as they give rise to a patterned, embryo-like structure in absence of species-specific external (micro-)environments and associated cues.\smallskip

\begin{figure}[H]
	\centering
	\includegraphics[width=3.2in]{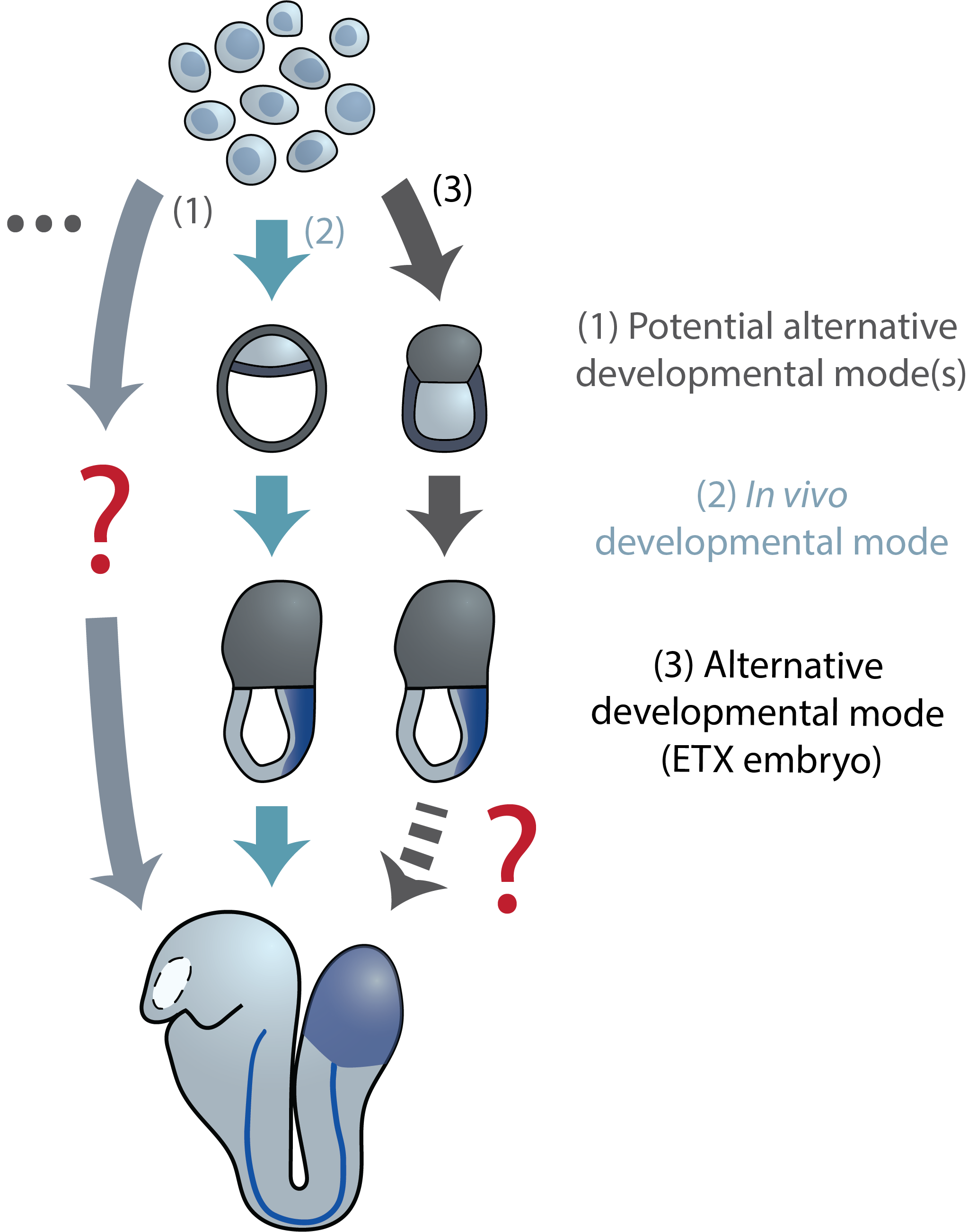}
	\caption{{\fignumfont Diverse developmental trajectories accessible by ESCs and ESC-LPs in the presence and absence of extraembryonic inputs.} Compared to the native embryo, (re-) aggregation of ESC-LPs reveals potentially alternative developmental modes to the same body plan. Evidence for this has been found in Nematostella reaggregates and mGastruloids\green{\cite{kirillova2018,hashmi2020,vianello2020}}. ETX embryos\green{\cite{sozen2018,zhang2019_3}} from mESCs, TSCs and XEN cells could point towards the existence of such an alternative mode, as they form structures resembling the actual gastrulating mouse embryo without closely mimicking blastocyst morphology (stages E3.5-4.5) earlier during their development. Since ETX embryos do not progress beyond the equivalent of stage E7.0-7.5, this alternative developmental mode remains a partial one.
		\label{Fig2}}
\end{figure}

It is noteworthy that this can indeed differ from the mode employed by the native embryo that uses a subset of (morphogenetic) trajectories available to the highly developmentally plastic embryonic cells \red{(Fig.\ \ref{Fig2})}.\smallskip
	
This plasticity is, for instance, exemplified by re-aggregates from \textit{Nematostella}, in which, during germ layer specification, trajectories of cells resemble those of cnidarians, such as hydrozoans. Yet, the end result is a functional animal, alike to one formed through normal \textit{Nematostella} embryogenesis\green{\cite{kirillova2018}}. In case of mGastruloids, cells expressing endodermal markers appear to be specified in a manner distinct to the embryo, as they arise dispersed and subsequently congregate without the requirement of an epithelial-to-mesenchymal transition (EMT)\green{\cite{vianello2020,hashmi2020}}.\smallskip

Complementary to the idea of ESC(-LP)s unraveling a conserved developmental mode under minimal culture conditions, is the observation that cells on a separate, advanced differentiation path can be coaxed to follow such basal trajectories when presented with the right factors. In re-aggregates from \textit{Nematostella} aboral halves, prior injection of Wnts is required to prompt the cells to develop into a functional animal\green{\cite{kirillova2018}}. Conversely, in gastruloids from induced pluripotent stem cell (iPSC) lines, AP patterning dynamics, including initial Bra/T polarization, are similar to their ESC counterparts\green{\cite{beccari2018}}.

%%%%%%%%%%%%%%%%%%%%%%%%%%%%%%%
%%%%%%%%%%%%%%%%%%%%%%%%%%%%%%%
\bigskip
\textbf{5. Significance of extraembryonic inputs in recapitulating embryo-like patterning} 
\\
Embryo-like structures also serve as a valuable tool to study the divergence of developmental trajectories in a systematic manner by exposing ESC(-LP)s to species-specific cues. These can be applied in manifold ways, for instance via adding early extraembryonic cells such as TSCs (trophoblast stem cells) or XEN (extraembryonic endoderm stem-) cells which in mammals give rise to tissues that closely interact with and instruct embryo patterning\green{\cite{arnold2009,langdon2011}}. Further options comprise the timed and localized application of signaling molecules to mirror endogenous inputs and the use of ECM mimics, such as hydro- or matrigel \green{\cite{hughes2010,caliari2016}} to generate 2D or 3D scaffolds and hence cues through mechanochemical interactions. \smallskip 

Indeed, several recently described systems which implement such principles are allowing us to dive deeper into the question of how embryonic cells can generate a richer diversity of cell states and embryo-like morphology in the presence of aforementioned external cues.\medskip

\noindent\textbf{Mouse - Gastruloids}
\\ By expanding on culture conditions of mGastruloids, further research efforts have coaxed these into recapitulating further aspects of native embryo morphology. For example, a recent work reports the application of matrigel to generate somite-like structutures in conjunction with respective gene oscillation during mGastruloid elongation\green{\cite{vandenbrink2020}}. A parallel study adds that Wnt inhibition during matrigel embedding further promotes anterior segment formation and improves physical separation of somites\green{\cite{veenvliet2020}}.\smallskip

Upon administration of cardiogenic factors FGF, VEGF and ascorbic acid, mGastruloids reproducibly recapitulate cardiogenesis at their anterior end, giving rise a vascular-like network, first and second heart fields as well as ultimately a beating structure, resembling an actual embryonic heart\green{\cite{rossi2019}}. Moreover, mGastruloids generated not only from mESCs but also from XEN cells exhibit increased cell type diversity and develop neural-tube-like structures.\green{\cite{berenger-currias2020}}.

\noindent\textbf{Mouse - Blastoids}
\\
Blastoids, made from aggregated mESCs and TSCs, illustrate how the two cell types influence each other to form structures closely resembling mouse E3.5 blastocysts (morphologically and transcriptionally) with an outer trophoectoderm layer encapsulating the blastocoel, a fluid filled cavity, as well as the epiblast\green{\cite{rivron2018}}. It appears that, not only do the TSCs guide the ESCs along their native developmental trajectory, the ESCs in turn maintain proliferation and self-renewal of TSCs as well as the trophoblast epithelial morphogenesis.\smallskip 

Although blastoids will not form bona fide embryos, they are capable of implanting \textit{in vivo}, thereby constituting a potentially useful model system for studying mouse pre-gastrulation and implantation patterning dynamics. Other works also demonstrate the assembly of similar systems\green{\cite{sozen2019,kime2019}}. In particular, EPS-blastoids are generated from extended pluripotent stem cells (EPSCs), pluripotent SCs cultured under conditions which enable both embryonic and extraembryonic developmental potential, as well as TSCs.\smallskip

These generate a primitive endoderm (PE)-like layer which, unlike normal blastoids, gives rise to cells expressing parietal endoderm markers. Moreover, under specific culture conditions, a subset of EPS-blastoids proceed to develop into a post-implantation embryo-like morphology, similar to ETX embryos and E5.0-E5.5 cultured mouse embryos. The majority of these advanced EPS-Blastroids also form an outer visceral endoderm (VE)-like layer.\medskip

\noindent\textbf{Mouse - ETS and ETX embryos}
\\
Single mESCs and TSCs, when grown together in a 3D Matrigel scaffold, mutually cooperate to assemble into structures, termed ETS embryos, highly reminiscent of the actual mouse embryo from E5.0 to 6.5\green{\cite{harrison2017}}. They recapitulate epiblast and trophoblast lumenogenesis, pro-amniotic cavity formation as well as mesoderm and primordial germ cell (PGC) induction.\smallskip 

Whereas in gastruloids the developing Bra/T$ ^{+} $ pole or posterior seems to have no directional preference in the spherical aggregate, in ETS embryos, on the other hand, the presence of TSCs spatially confines the emergence of the this Bra/T domain at the boundary between embryonic and extraembryonic compartments, akin to the natural embryo. This role of the extraembryonic tissues is in congruence with the \textit{in vivo} situation, where they provide not only the right signalling environment but also the crucial geometric cues that confer robustness to a developing embryo\green{\cite{arnold2009,langdon2011,christodoulou2019}}.\smallskip

This system was further complemented by adding a third cell type, XEN cells, which augment morphological similarity to \textit{in vivo} mouse embryos by adding a visceral endoderm (VE)-like layer\green{\cite{sozen2018,zhang2019_3}}.\ Moreover, resulting ETX embryos display EMT, and subsequently meso- as well as endodermal marker expression, thus mimicking mouse embryo shape and morphogenesis until E7.0-7.5. It must be noted that, while such ETX embryos could initiate implantation responses in mouse uteri upon transplantation, they do not progress beyond this stage.\medskip

\noindent\textbf{Human - Amniotic-sac-like embryoids}
\\ 
By employing a microfluidic device, researchers were able to recapitulate post-implantation embryo development prior to and during early gastrulation, encompassing epiblast luminogenesis, formation of the bipolar embryonic sac and PGC as well as primitive streak cell specification\green{\cite{zheng2019}}. To achieve this, single hESCs were grown on gel pockets within a customized three channel device, enabling cell loading and consecutive medium switching.\smallskip

Following lumenogenesis, dorsal amniotic ectoderm-like cells were induced via localized application of BMP4. Ventral epiblast-like cells give rise to subpopulations expressing primitive streak markers such as BRA/T and PGC-associated genes. After 2-3 days in culture, cysts collapse due to emigration of ventral cells.\smallskip

This example again nicely illustrates, in a different species than mouse, how embryo-like morphology can be recapitulated through mimicking of an extraembryonic environment which evidently provides structural instructions.

%Collectively, the above studies in mouse and human systems hence suggest that this can be achieved via adding extraembryonic stem cell types which self-organize into respective tissues or through recapitulating (aspects of) the biomechanical and signaling input provided by these structures.\vspace*{0.3cm}
\medskip
%note:smallskip or medskip here don't work well)
\noindent\textbf{Mouse and human - Micropatterns} 
\\
Another well established proxy for human gastrulation are the disc-shaped micropatterns used to spatially confine hESC colonies\green{\cite{peerani2007,bauwens2008}}. Treatment with BMP4 elicits germ layer organization and associated marker gene expression reminiscent of early embryos\green{\cite{warmflash2014}}. Notably, in conventional culture conditions, without a boundary constraint imposed by micropatterns, spatial patterns of differentiation differ drastically between neighboring colonies despite exposure to BMP4. This argues for an important role of precise control of size and geometry, constricting the hESCs along a developmental path to generate patterns in concentric radial domains with trophoectoderm-, mesendoderm and ectoderm-like fates separated along the radial axis.\smallskip 

Micropatterns have proven to be a useful tool for probing mechanisms of cell fate specification, for instance by modelling neurulation\green{\cite{etoc2016,nemashkalo2017,tewary2017,haremaki2019}}. Furthermore, a related study succeeded in inducing PS-like marker gene expression and morphology as well as an organizer-like cell population\green{\cite{martyn2018}}. However, while the micropattern’s disc-like shape resembles that of the human embryo, their radially symmetric gene expression patterns are unlike the axial organization in actual animals, which may be due to the structural constraints imposed by the micropattern itself. Micropatterns with radial arrangement of germ layers have also been generated from mouse epiblast-like cells (EpiLCs)\green{\cite{morgani2018}} which are more similar to hESCs in transcriptional state and culture conditions than to mESCs\green{\cite{tesar2007,rossant2017}}.\smallskip

In this light, human and mouse micropatterns may therefore reveal another developmental mode, different to the one displayed by gastruloid-like systems. While, in the latter case, cells self-organize into similar 3D structures in absence of geometric constraints, in the former case, micropatterns facilitate ESCs to converge onto a mode resulting in the formation of radially patterned germ layers.

%%%%%%%%%%%%%%%%%%%%%%%%%%%%%%%
%%%%%%%%%%%%%%%%%%%%%%%%%%%%%%%
\bigskip

\textbf{6. Outlook}
\\
In conjunction with the observed diversity in (pre-) gastrulation embryo shapes, comparison of \textit{in vitro} and \textit{ex vivo} systems to their \textit{in vivo} counterparts may point towards an underlying developmental morphospace spanning a plethora of available trajectories\green{\cite{mitteroecker2009}}. Since embryos of a given species generally develop under the same specific inputs, including geometry and extraembryonic environment, ESCs are developmentally biased and such alternative trajectories are revealed only when cells are removed from their native context and challenged by different external conditions \red{(Fig.\ \ref{Fig3})}. Furthermore, as illustrated above, ESC populations appear to possess an inherent capability to at least form a primary axis through a conserved developmental mode, which surfaces when ESC(-LP)s are grown in the most minimal conditions.\smallskip

When comparing systems such as gastruloids with blastoids and ETS/ETX embryos it becomes apparent that, while the latter (mostly) faithfully recapitulate embryo geometry and morphogenesis, they are not able to develop beyond the \textit{in vivo} equivalent of embryonic stage E7.0. On the other hand, gastruloids do not look like actual embryos and undergo a more abstracted morphogenesis, yet they are capable of reaching the partial (transcriptional) equivalent of E9.0 (in case of the mESC-based system).\smallskip

These observations raise an interesting point: It seems that by employing additional early embryonic cell types (i.e. mimicking a set of \textit{in vivo} cues) resulting embryo-shaped artificial systems remain more constrained in their developmental potential, likely because once having assumed an embryo-like developmental mode, they would require further, precise spatially and temporally allocated inputs, just like the \textit{in vivo} counterpart, to facilitate successive morphogenetic events. Gastruloid-like aggregates from pure ESC or ESC-LPs, on the other hand, might represent a more unconstrained system, which is what enables them to develop beyond the equivalent of initial gastrulation, and therefore rather unveil the extent of inherent self-organizational capabilities of the source cells.\smallskip

In summary, the wide range of trajectories displayed by ESCs and ESC-LPs \textit{in vitro} converge onto an aforementioned conserved developmental mode under minimal conditions and collapse partly onto their native trajectories (at least for the mammalian systems) when provided with external inputs to recapitulate species-specific external environments (mechanical and chemical), facilitating increased morphological resemblance to the native embryo.\smallskip

Characterizing this fundamental mode may perhaps ultimately provide insights into the origin and evolution of animal body plans as well as how the complexity of animal shapes during peri-gastrulation development may have arisen via distinct extraembryonic factors. In order to properly address this from a comprehensive evolutionarily perspective, however, more phylum coverage is required, dictating the need for novel minimal \textit{in vitro} embryo-like systems from, for instance, diploblasts and protostomes.\smallskip

We note that there are established concepts related to conservation of body plans and gene expression across species: The developmental hourglass model delineates a phylotypic period at mid-embryogenesis during which common anatomical features (the basic body plan) for each respective phylum are established\green{\cite{duboule1994,raff2012,rittmeyer2012}}. In turn, this period is developmentally preceded and followed by phases of increased morphological divergence, thereby shaping a developmental ``hourglass''\green{\cite{irie2014}}.\smallskip

A related term is ``mid-developmental transition'' (MDT), referring to a transition period between early and late stages of conserved gene expression which is characterized by phylum-specific activity of gene networks. This phase, overlapping with the phylotypic period in previously studied animals, has been identified from sponges to chordates and may therefore serve to define a phylum as a group of species that, during the MDT, exhibit gene expression which is convergent among themselves but divergent to other species\green{\cite{levin2016}}.\smallskip

The conserved developmental mode that we propose here, on the other hand, applies to ESCs across phyla (albeit we acknowledge that this inference may be erroneous due to limited taxon coverage of existing minimal \textit{in vitro} systems). Moreover, this conserved mode entails the establishment of primary axial identity from an initially homogeneous or disorganized cell population, whereas the MDT and phylotypic period span the organogenesis phase of embryonic development, at which point the respective body axes are already specified.\smallskip

%Hence, the conserved developmental mode may help to explain pre- and peri- gastrulating embryo shape diversity prior to the mid-developmental transition, in so far as ESCs across species harbour substantial inherent developmental potential, enabling them to reach the onset of the MDT. This potential could, in turn, imply less constraints on early embryo morphodynamics and transcription by acting as a buffer to numerous environmental and life history-related challenges which changes as initial axis specification completes, dictating the need for more distinctly coordinated gene expression to generate phylum-specific body plans.

Regarding the significance of extraembryonic environments it could be argued that, from the multitude of ways that pre-gastrulating embryos have at their disposal to establish their initial body plan, refinement of a species-specific developmental mode as well as corresponding embryo morphology over the course of evolution ensures robustness (or canalization) and thus ultimately fitness\green{\cite{siegal2002,felix2008,wagner2008,foote1997,deline2018,melzer2016}}: Those developmental trajectories that worked most robustly in the specific reproductive niche were further specified to maximize fitness.\smallskip
	
While this hypothesis remains unverified in metazoan embryos at large, a recent study finds that diversification of insect egg size and shape, traits which vary greatly across species, are driven by shifts in the respective oviposition microenvironments, that is, where the eggs are laid\green{\cite{church2019}}.\smallskip 

Such external or extraembryonic environments and their complexity vary greatly across species and constitute a substantial challenge to replicate \textit{in vitro}. It is natural to speculate that this could be one of the reasons why it is more straightforward to generate a functional \textit{Nematostella} or Starfish larva as opposed to a fish, mouse or human from isolated ESC-like cells.\smallskip

Nonetheless, comparing these different kinds of embryo-like structures can therefore provide clues as to which parts of embryonic development or, more precisely, axis formation and associated morphogenesis are inherent to ESCs and which parts strictly require extraembryonic input. It has to be mentioned that with advances in biomaterials and biotechnology in general it should be possible to mimic such extraembryonic environments more closely in the future, and, in case of mammals, entirely substitute extraembryonic cell types\green{\cite{gritti2020}}.\smallskip 

Taken together, investigation of initial patterning events in artificial embryo-like systems will help to assess if and to which degree aforementioned conserved developmental modes exist. Hence, integration and comparison of such studies with data from actual embryos should aid to more concisely delineate underlying mechanisms of self-organized primary axis and subsequent initial body plan specification in cell ensembles as well as to elucidate early (pre- and peri-gastrulation) embryo morphology diversification during evolution.

%
%
%
% \bigskip
% 
%\begin{references}
\bibliographystyle{science}
\sffamily
\begin{footnotesize}
\bibliography{SB_evo}
\end{footnotesize}

%\end{references}
\begin{acknow}
We thank David Oriola, Nicola Gritti, Miki Ebisuya, Maria Costanzo and Jia Le Lim for their comments on the manuscript. 
 \end{acknow}

%\begin{conflict}
%The authors have no competing interests. 
% \end{conflict}

\begin{funding}
K.A. and V.T. were supported by the European Molecular Biology Laboratory (EMBL) Barcelona. 
 \end{funding}

%\begin{suppinfo}
%Materials and methods, Supplementary figures and Supplementary movies. 
%\end{suppinfo}
%\medskip
%
%
%
\end{document}